\documentclass[aps,pra,superscriptaddress,twocolumn]{revtex4-2}
\usepackage{amssymb}
\usepackage{graphicx}
\usepackage{amsmath}
\usepackage[colorlinks=true,linkcolor=blue,citecolor=blue,urlcolor=blue]{hyperref}

\begin{document}

\title{Universal Dynamic Scaling and Contact Dynamics in Quenched Quantum Gases}

	\author{Jia-nan Cui}
	\affiliation{School of Science, Beijing University of Posts and Telecommunications, Beijing 100876, China}
	
	\author{Zhengqiang Zhou}
	\affiliation{School of Science, Beijing University of Posts and Telecommunications, Beijing 100876, China}

\author{Mingyuan Sun}
\email{mingyuansun@bupt.edu.cn}
\affiliation{School of Science, Beijing University of Posts and Telecommunications, Beijing 100876, China}
\affiliation{State Key Lab of Information Photonics and Optical Communications, Beijing University of Posts and Telecommunications, Beijing 100876, China}

\date{\today}

\begin{abstract}

Recently universal dynamic scaling is observed in several systems, which exhibit a spatiotemporal self-similar scaling behavior, analogous to the spatial scaling near phase transition. The latter one arises from the emergent continuous scaling symmetry. Motivated by this, we investigate the possible relation between the scaling dynamics and continuous scaling symmetry in this paper. We derive a theorem that the scaling invariance of the quenched Hamiltonian and the initial density matrix can lead to the universal dynamic scaling. It is further demonstrated both in a two-body system analytically and in a many-body system numerically. For the latter one, we calculate the dynamics of quantum gases quenched from the zero interaction to a finite interaction via the non-equilibrium high-temperature virial expansion. A dynamic scaling of the momentum distribution appears in certain momentum-time windows at unitarity as well as in the weak interacting limit. Remarkably, this universal scaling dynamics persists approximately with smaller scaling exponents even if the scaling symmetry is fairly broken. Our findings may offer a new perspective to interpret the related experiments. We also study the Contact dynamics in the BEC-BCS crossover. Surprisingly, the half-way time displays a maximum near unitarity while some damping oscillations occur on the BEC side due to the dimer state, which can be used to detect possible two-body bound states in experiments.       

\end{abstract}

\maketitle

\section{Introduction}

Quench dynamics refers to the evolution of a closed system after a sudden change of either the initial state or the Hamiltonian. Given its simple operation in experiments, it becomes one of the most studied non-equilibrium dynamics, especially in ultracold atomic gases~\cite{quench_exp1,quench_exp2,quench_exp3,quench_exp4,quench_exp5,Zoran,
quench_exp6,quench_exp7,quench_exp8}. The advance in experiments further prompts extensive theoretical studies~\cite{theory1,theory2,theory3,theory4,theory5,theory6,theory7,fop1,theory8,
theory9,theory10,theory11,theory12,Gao,Parish,Colussi,Bougas1,Sun,Colussi1,Bougas2,Colussi2,
Enss,fop2,fop3}. For instance, the Cambridge group observed several universal phenomena in a three-dimensional (3D) Bose gas quenched from noninteracting to unitarity, such as the exponential form of the momentum distribution at the final steady state~\cite{quench_exp3}. Later, the low temperature condensed part and the thermal gas part of the data are respectively explained by using the time-dependent Bogoliubov theory~\cite{Gao,Parish} and the non-equilibrium high-temperature virial expansion~\cite{Sun}. On the other hand, some characteristics of the quenched Hamiltonian can lead to novel dynamical phenomena~\cite{MBL,linking_th,linking_exp1,linking_exp2,Gao_Shi}. For instance, the Chern number of a topological insulator can be revealed by the linking number in quench dynamics~\cite{linking_th,linking_exp1,linking_exp2}. The discrete scaling symmetry of the quenched Hamiltonian may give rise to a dynamical fractal~\cite{Gao_Shi}. 

Recently, universal scaling dynamics is observed in quenched Bose gases~\cite{quench_exp4,quench_exp5,Zoran,quench_exp6,quench_exp8}. The scaling property of spin correlations is investigated both in a quasi-1D~\cite{quench_exp4} and 2D~\cite{quench_exp8} spinor Bose gas, while other works demonstrate the dynamic scaling behavior of the momentum distribution in 1D~\cite{quench_exp5}, 2D~\cite{quench_exp6} and 3D~\cite{Zoran} Bose gases respectively. Specifically, the evolution of the momentum distribution exhibits a spatiotemporal self-similar scaling feature in certain momentum-time windows, which is analogous to the spatial scaling around the critical point of phase transition~\cite{Huang,Sachdev}. The latter arises from the continuous scaling invariance of the system emerged at the critical point, where the correlation length is divergent. The scaling dynamics has been studied in various systems by a number of theoretical works~\cite{dyt1,dyt2,dyt3,dyt4,dyt5,dyt6,dyt7,dyt8,dyt9,dyt10,dyt11,dyt12,Berges}, mainly in the framework of non-thermal fixed points. However, there are some disagreements between the theory and the experiment, for example, the scaling exponent~\cite{quench_exp5,Zoran}. On the other hand, the possible relation between the scaling symmetry and the dynamic scaling has not been discussed yet. If the Hamiltonian $\hat{H}(\mathcal{P})$ has a continuous scaling symmetry, i.e, $\hat{H}(\zeta\mathcal{P})=\zeta^\beta\hat{H}(\mathcal{P})$ with a scaling exponent $\beta$ ($\zeta$ is a positive constant), it can play an important role in quench dynamics through the time-evolution operator $e^{-it\hat{H}}$. 

In this letter we investigate this problem by deriving a theorem that connects the continuous scaling symmetry of the quenched Hamiltonian and the initial density matrix to the dynamic scaling of the momentum distribution. We further demonstrate its validity both in a two-body system and in a many-body system. Surprisingly, even if the scaling symmetry required by the theorem is fairly broken, an approximate scaling dynamics can still occur with modified scaling exponents in certain momentum-time windows, which offers a new perspective to interpret the experimental observations.  At last, we calculate the Contact dynamics to investigate how the short-range correlations are built up in quenched quantum gases~\cite{theory2, theory5, theory8, Colussi}, since the Contact is a universal quantity to characterize the property of short-range interactions~\cite{Tan, Zhang}. Recently, the Contact dynamics also receives the experimental interest~\cite{quench_exp1, Contact_exp1, Contact_exp2, Contact_exp3}. 

\section{Universal dynamic scaling}

\textit{Theorem.} If the initial density matrix $\rho_0=\int d\kappa w_\kappa | \psi_\kappa \rangle  \langle  \psi_\kappa | $ and the quenched Hamiltonian $\hat{H}({\bf p_1},{\bf p_2},...)$ both have a continuous scaling symmetry, i.e.,
\[
  \begin{cases}
    w_{\zeta\kappa}=\zeta^\delta w_\kappa    \\
    \hat{H}(\zeta{\bf p_1},\zeta{\bf p_2},...)=\zeta^\gamma\hat{H}({\bf p_1},{\bf p_2},...).
  \end{cases}
\]
where, $\delta$ and $\gamma$ denote the scaling exponents. The momentum distribution $n({\bf k},t)$ will exhibit a dynamic scaling behavior in the momentum-time domain, which can be expressed as
\begin{equation}
    n({\bf k},t) = \tilde{t}^\alpha n(\tilde{t}^\beta {\bf k}, t_0)
    \label{nktscaling}
\end{equation}
with $\beta=1/\gamma$ and $\alpha=2d/\gamma$. Here, $\kappa$ represents the set of the quantum numbers for the eigenstate $|\psi_\kappa \rangle$. For example, for non-interacting quantum gases, $\kappa=\{\bf p_1, p_2,...\}$, and for an interacting two-body system with no bound states, $\kappa=\{\bf P, q \}$ with ${\bf P}$ and ${\bf q}$ being the total and relative momentum respectively.  $\zeta$ is a positive number (i.e., the scaling factor). $\tilde{t}=t/t_0$ and $d$ denotes the dimensionality of the system. The proof is displayed in the following.

For a given initial density matrix $\rho_0=\int d\kappa w_\kappa | \psi_\kappa\rangle  \langle  \psi_\kappa | $ at time $t=0$, the dynamics of the momentum distribution $n({\bf k},t)$ governed by the quenched Hamiltonian $\hat{H}$ at time $t>0$ can be expressed as
\begin{equation}
   n({\bf k},t)=\int d\kappa  w_\kappa  \langle  \psi_\kappa | e^{it\hat{H}}\hat{n}_{\bf k} e^{-it\hat{H}} | \psi_\kappa \rangle
    \label{nkt}
\end{equation}
For simplicity, we define a scaling operator $\hat{S}_\zeta$ as $\langle {\bf p_1},{\bf p_2},... | \hat{S}_\zeta | \phi_\kappa \rangle = \langle {\bf p_1}/\zeta,{\bf p_2}/\zeta, ... | \phi_\kappa \rangle \equiv \zeta^{Nd} \langle {\bf p_1},{\bf p_2},... | \phi_{\zeta\kappa} \rangle$, with $N$ being the total particle number (see the appendix).  By inserting the basis of the eigenstates of the quenched Hamiltonian $\hat{H}$, one can connect the momentum distribution $n({\bf k},t)$ with $n(\zeta{\bf k},t\zeta^{-\gamma})$, which is displayed as follows (see the appendix). 
\begin{equation}
  n({\bf k},t)=\zeta^{2d} n(\zeta {\bf k}, t\zeta^{-\gamma}) \equiv \tilde{t}^{\alpha} n(\tilde{t}^{\beta} {\bf k}, t_0)
   \label{nkt1}
\end{equation}
Note that $\delta=-Nd$ from the constraint $\int d\kappa  w_\kappa = 1$. Here, we have used the property that $|\phi_{\zeta\kappa}\rangle$ is also the eigenfunction of the Hamiltonian with the eigenvalue $\zeta^\gamma E_\kappa$~\cite{Felix}. In the following, we demonstrate this theorem by directly calculating the quench dynamics of the momentum distribution in a two-body system as well as in a many-body system via the framework of the high-temperature virial expansion.

\textit{Two-body system.}
We first consider a two-body system with equal mass ($m$) and an s-wave interaction between them. Essentially, it can be transformed into a one-body problem and the Hamiltonian can be written as $\hat{H}=-\nabla^2/m+\hat{V}({\bf r})$ in the relative coordinate. We take the reduced Planck constant and Boltzmann constant equal to unity for simplicity in this paper. If the system is quenched to a finite scattering length $a_s$ at $t=0$ with an initial state $|{\bf k}={\bf 0}\rangle$, the momentum distribution $n({\bf k},t)$ for finite momenta (${\bf k}>0$) after the quench can be expressed as
\begin{equation}
   n({\bf k},t) =|\frac{1}{2\pi i}\int_{-\infty}^{+\infty} d\omega \frac{t_2(s) e^{-i \omega t}}{s(s-\varepsilon_{\bf k})} |^2 
      \label{nkt2b}
\end{equation}
with the two-body scattering matrix
\begin{equation}
   t_2(s)=\frac{4\pi}{m}\frac{1}{a_s^{-1}-\sqrt{-ms}}
\end{equation}
Here, $s=\omega+i0^+$ and $\varepsilon_{\bf k}={\bf k}^2/m$. For any finite scattering length, the dynamics of the momentum distribution does not follow Eq.~(\ref{nktscaling}), due to the breakdown of the scale invariance. However, in the unitary limit ($a_s\rightarrow\infty$), the scale invariance recovers and Eq.~(\ref{nkt2b}) can be simplified to
\begin{equation}
   n({\bf k},t)=\frac{64\pi t}{ m k^4}|F(\frac{1-i}{\sqrt{2 m}} k\sqrt{t})|^2
\end{equation}
where, the function $F(z)=1-\frac{\sqrt{\pi}}{2}\frac{{\rm erf}(z)}{z}e^{z^2}$ with ${\rm erf}(z)$ being the error function. One can see that it indeed satisfies Eq.~(\ref{nktscaling}) and exhibits the dynamic scaling with the scaling exponents $\beta=1/2$ and $\alpha=3$ as the theorem says.

If one considers the weak interacting limit, $a_s\rightarrow 0$, to the lowest order, the momentum distribution can be written as
\begin{equation}
   n({\bf k},t)=\frac{32\pi^2 a_s^2 }{k^4}[1-{\rm cos}(\frac{k^2t}{m})]
\end{equation}
Remarkably, it also displays the dynamic scaling behavior but with the scaling exponents $\beta=1/2$ and $\alpha=2$. Because the scattering length is much smaller than other relevant length scales, the system remains scale-invariant approximately and our theorem is still valid in such cases. But the scaling exponents can depend on the specific approximation, which is also indicated in Eq.~(\ref{nkt1}).

\textit{Many-body system.} For a many-body system, the particle density (or the distance between nearest particles) introduces a new length scale, which generally prevents the system from owning the continuous scaling symmetry, such as the non-interacting or unitary Fermi gas. However, in the low-density limit or the high-energy regime, the influence of this length scale can be negligible and the many-body system may also exhibit scale invariant, similar to the two-body system. In the following, we employ the high-temperature virial expansion to demonstrate the dynamic scaling, where our theorem is applicable.
\begin{widetext}

\begin{figure}[t] 
    \centering
    \includegraphics[width=0.95\textwidth]{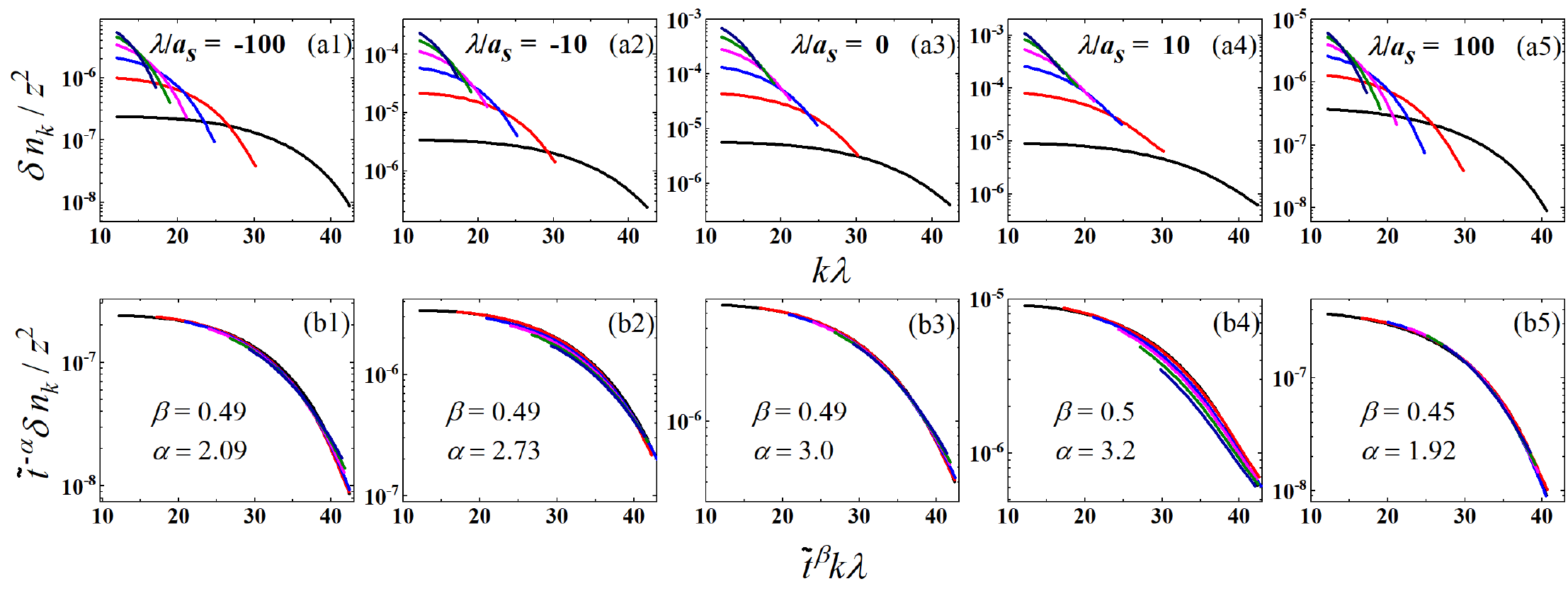}
    \caption{The momentum distribution $\delta n_{\bf k}$ at the large momentum region for various times $t/t_\lambda$ = 0.02(black), 0.04(red), 0.06(blue), 0.08(pink), 0.10(green), 0.12(navy) in the quench dynamics of quantum gases. Here, $\lambda=\sqrt{2\pi/(mT)}$ is the thermal wavelength, while $t_\lambda=1/T$. $\tilde{t}=t/t_0$ with $t_0/t_\lambda=0.02$. The results in the BEC-BCS crossover are displayed with the scattering length labelled in the first row. In the second row, the corresponding scalings of both the horizontal and vertical axes are taken via Eq.~(\ref{nktscaling}), to demonstrate whether there exists a dynamic scaling behavior. The scaling exponents $\alpha$ and $\beta$ are  shown in the corresponding graphs.  }
     \label{fig1}
\end{figure}

\end{widetext}

If one considers that a quantum gas is quenched from a non-interacting Hamiltonian $\hat{H}_0$ with an equilibrium state to an interacting Hamiltonian $\hat{H}$ at time $t=0$, the evolution of the momentum distribution $n({\bf k},t)$ for $t>0$ can be expressed as  
\begin{equation}
   n({\bf k},t)=\frac{\text{Tr} [ e^{-\beta(\hat{H}_0-\mu \hat{N})}e^{it\hat{H}}\hat{n}_{\bf k} e^{-it\hat{H}} ]}{\text{Tr}[e^{-\beta(\hat{H}_0-\mu \hat{N})} ]}
    \label{nkt}
\end{equation}
Here $\beta=1/(k_BT)$ with $T$ being the temperature. $\mu$ and $N$ are the chemical potential and the total number of particles, respectively. At high temperature, it can be expanded in powers of the fugacity $z=e^{\beta\mu}\ll 1$ in the framework of virial expansion~\cite{Sun}. In the following, we will consider the change of the momentum distribution $\delta n({\bf k},t)=n({\bf k},t)-n({\bf k},0)$ for simplicity, since it represents the property of the quenched Hamiltonian $\hat{H}$. To the second order of $z$, it can be expressed as 
\begin{align}
  \delta n({\bf k},t) =& 2z^2 \sum_{\bf k_1, k_2} f({\bf k_1}) f({\bf k_2}) \nonumber \\ 
      &\{ 2{\rm Re}[g({\bf k_1},{\bf k_2},t)]\delta_{{\bf k_1},{\bf k}} +|g({\bf k_1},{\bf k_2},t) |^2 \}
      \label{deltnkt}
\end{align}
with
\begin{equation}
  g({\bf k_1},{\bf k_2},t)=\frac{1}{2\pi i}\int_{-\infty}^{+\infty} d\omega \frac{t_2(s) e^{-i t(\omega-\varepsilon_{\bf q_1})}}{(s-\varepsilon_{\bf q_1})(s-\varepsilon_{\bf q_2})}
  \label{gt}
\end{equation}
Here, $s=\omega+i0^+$, ${\bf q}_1=({\bf k}_1-{\bf k}_2)/2$, ${\bf q}_2={\bf k}-({\bf k}_1+{\bf k}_2)/2$, $\varepsilon_{\bf q}={\bf q}^2/m$ and $f({\bf k_1})=e^{-2\beta\varepsilon_{\bf k_1}}$. One can calculate Eq.~(\ref{deltnkt}) numerically to obtain the dynamics of the momentum distribution.

The initial distribution function $f({\bf k_1})=e^{-2\beta\varepsilon_{\bf k_1}}$ does not have a scaling symmetry. However, if one considers the large momentum region $k\lambda \gg 1$ and $t/t_\lambda \ll 1$($t_\lambda=1/T$), this Gaussian function can be regarded as a Dirac delta function approximately and the scaling symmetry recovers. Therefore, according to our theorem, a scaling dynamics can occur at unitarity~\cite{footnote1} and indeed this is ture as shown in Fig.~\ref{fig1}. Moreover, we calculated the results in the whole BEC-BCS crossover. The finite scattering length breaks the scaling symmetry of the Hamiltonian and the dynamic scaling disappears. Nevertheless, in the weak interacting limit, i.e, $a_s\rightarrow 0$, the scattering length is much smaller than all other relevant length scales in the system and the scaling symmetry recovers approximately, which further leads to the recurring scaling dynamics. Note that to the lowest order (i.e., the order of $a_s^2$), $\beta=1/2$ and $\alpha=2$ due to the prefactor $a_s^2$ in this approximation, which does not follow the theorem quantitatively. All the numerical results agree well with the above analysis based on the theorem. Our results imply that the scaling dynamics can occur in a specific momentum-time window, as long as the initial state and the Hamiltonian are scale-invariant approximately in this domain.

Actually, if one takes the initial distribution function as $f({\bf k_1})=\delta({\bf k_1})$ mathematically, Eq.~(\ref{deltnkt}) can be simplified to
\begin{equation}
    \delta n({\bf k},t) = 2z^2 |g({\bf 0},{\bf 0},t) |^2 
\end{equation}
which is the same with the result of the two-body system (see Eq.~(\ref{nkt2b})), except the prefactor $2z^2$. Thus, it is straightforward to show that $\delta n({\bf k},t) = \tilde{t}^3\delta n(\tilde{t}^{1/2}{\bf k},t_0)$ for all finite momenta at unitarity and $\delta n({\bf k},t) = \tilde{t}^2\delta n(\tilde{t}^{1/2}{\bf k},t_0)$ as $a_s\rightarrow 0$ (to the order of $a_s^2$), which are in excellent agreement with our numerical results (see Fig.~\ref{fig1}).

\begin{figure}[t] 
    \centering
    \includegraphics[width=0.49\textwidth]{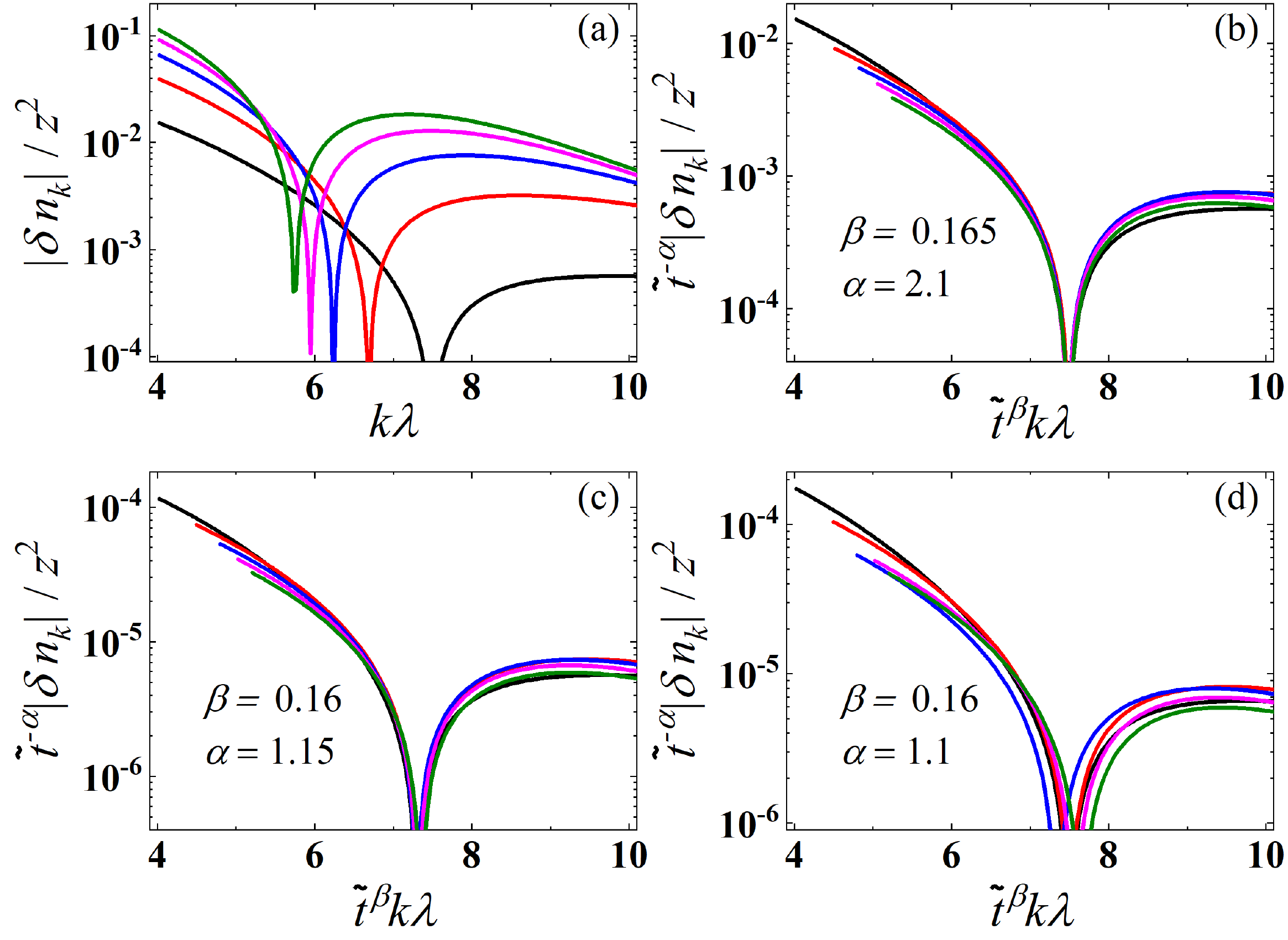}
    \caption{The quench dynamics of the momentum distribution $|\delta n_{\bf k}|$ at finite momentum region for various times $t/t_\lambda$=  = 0.1(black), 0.2(red), 0.3(blue), 0.4(pink), 0.5(green). The results for the unitarity are plotted with unscaled axes (a) and scaled axes (b), while they are only demonstrated in scaled axes for $\lambda/a_s=-100$ (c) and  $\lambda/a_s=100$ (d). According to the theorem, they are not expected to display any strict dynamic scaling behavior, due to the breakdown of the scale invariance of the initial density matrix. However, an approximate scaling dynamics can still be seen in all three cases with smaller scaling exponents $\alpha$ and $\beta$ displayed in the corresponding graphs.   }
     \label{fig2}
\end{figure}

\textit{Approximate dynamic scaling.}
Surprisingly, even if the scale-invariant condition is not strictly satisfied, an approximate scaling dynamics can still occur in certain momentum-time windows as shown in Fig.~\ref{fig2}. If one considers the momentum region $k\lambda \in [4,10]$, the strict dynamic scaling is not expected given the missing scaling invariance of the initial density matrix. However, there exists an approximate dynamic scaling with modified scaling exponents, which may arise from the approximate scaling behaviors of the initial density matrix (or the main contributing part) with different scaling exponents in certain momentum regions. 

On the one hand, it demonstrates that the dynamic scaling can be robust against the proper breakdown of the continuous scaling symmetry for either the initial density matrix or the Hamiltonian. On the other hand, it could be closely related to the recent experimental observations~\cite{quench_exp5, Zoran}, since the experimental condition does not strictly satisfy the scaling symmetry either and the scaling exponents vary significantly in different systems. Here, we would like to emphasize that there are two key features, on which both the experiment and our theory agree well with each other qualitatively. One is that both the scaling exponents $\beta$ and $\alpha$ are smaller than that of the strict dynamic scaling in the theorem. The other is the scaling exponents depend on the chosen momentum-time windows. Therefore, our results provide a new perspective to interpret the experimental observations.

\section{Contact dynamics}

\begin{figure}[t] 
    \centering
    \includegraphics[width=0.4\textwidth]{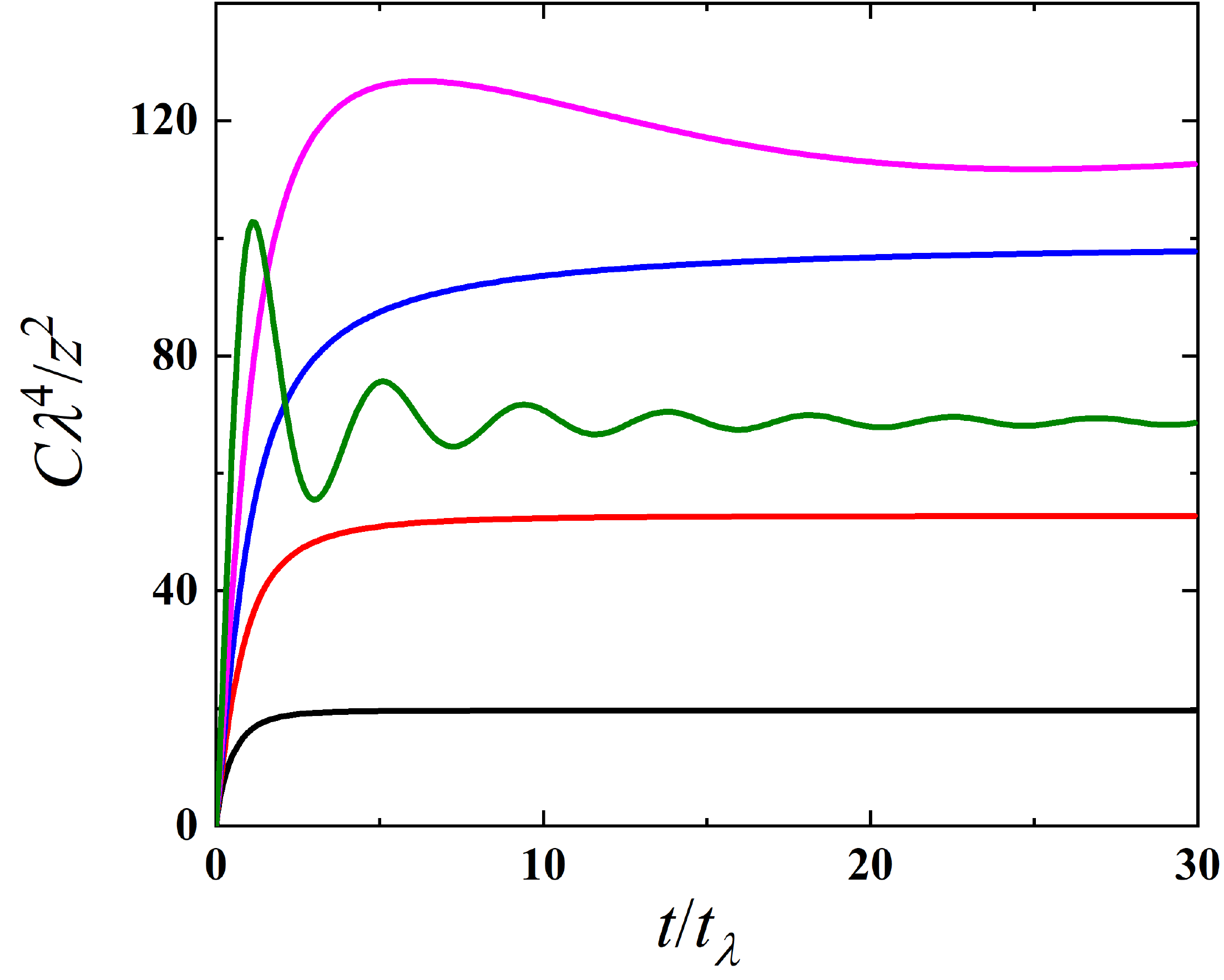}
    \caption{The Contact dynamics after the quench for $\lambda/a_s=$-3(black),-1(red),0(blue),1(pink),3(green). They all start from zero and approach to a steady value at long times. On the BEC side, it exhibits damping oscillations.  }
     \label{fig3}
\end{figure}

In the following, we further study universal features of the Contact dynamics in quenched quantum gases, which correspond to the large-momentum limit. In the framework of high-temperature virial expansion, it can be directly obtained from the momentum distribution, through the relation $C(t)= \lim_{k\rightarrow \infty} k^4 \delta n({\bf k},t)$. To the second order of $z$, the Contact dynamics in the BEC-BCS crossover are displayed in Fig.~\ref{fig3}. Owing to the initial non-interacting equilibrium state, they all start from zero and gradually approach to a steady state at long evolution times. At unitarity, one can derive an analytic form, expressed as
\begin{equation}
  C(a_s\rightarrow \infty)= \frac{64z^2}{\lambda^4} {\rm arctan}(\frac{t}{t_\lambda})
\end{equation}

In the whole BEC-BCS crossover, as $t\rightarrow 0$, the Contact can be expanded in powers of $t$. To the lowest order, it can be written as 
\begin{equation}
  C(t\rightarrow 0)= \frac{64z^2}{\lambda^4 t_\lambda} t
\end{equation}
which is independent of the scattering length and consistent with Ref.~\cite{Qi}'s result. Actually in this limit, one can obtain the momentum distribution $\delta n({\bf k},t)\approx (64z^2/\lambda^4 t_\lambda)t/k^4$ for large momenta. As expected, it satisfies the dynamic scaling relation (see Eq.~(\ref{nktscaling})), because the scattering length is much larger than other relevant length scales in this regime and the system can be regarded as being scale-invariant. On the BEC side, there exist some damping oscillations, which are associated with the dimer state. Roughly speaking, they result from the coupling between the dimer state and the scattering states after the quench.  The decoherence of the scattering states further leads to the damping, due to different scattering energies.

\begin{figure}[t] 
    \centering
    \includegraphics[width=0.48\textwidth]{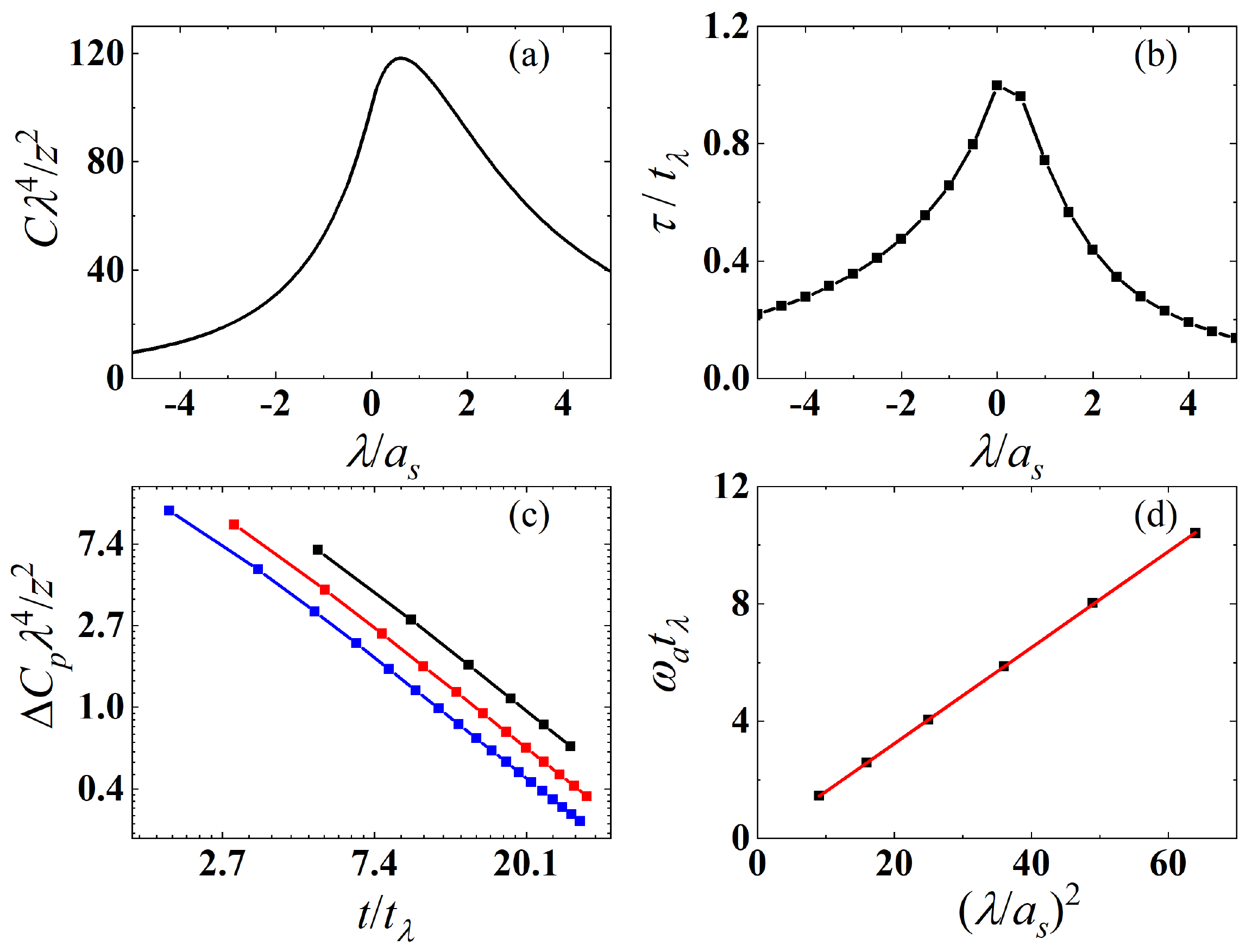}
    \caption{The Contact dynamics in the BEC-BCS crossover. (a) The Contact at infinite time ($t\rightarrow \infty$) in the whole BEC-BCS crossover, obtained from Eq.~(\ref{Contf}). (b)The half-way time $\tau$ as a function of the interaction. (c)The relative value at the peak $\Delta C_p=C_p-C(t\rightarrow\infty)$ for $\lambda/a_s=$3(black),4(red),5(blue), which obeys a power-law decay with the same power approximately. (d) The frequencies of oscillations at various interactions on the BEC side, with a linear fitting (red line).  }
     \label{fig4}
\end{figure}

For $t\rightarrow\infty$, one can obtain the Contact of the final steady state, which can be written analytically as
\begin{align}
  C(\infty)=& \frac{z^2}{\lambda^4} [ (32\pi-\frac{64 \lambda^2}{a_s^2}\Theta(a_s)) \nonumber \\
         & +\frac{16\sqrt{2}\lambda}{a_s}(\pi+\frac{2\lambda^2}{a_s^2}\Theta(a_s))e^{\frac{\lambda^2}{2\pi a_s^2}}{\rm erfc} (\frac{\lambda}{\sqrt{2\pi}|a_s|})  ] \nonumber \\
  \label{Contf}
\end{align}
Here, $\Theta(x)$ and erfc($x$) are respectively the unit step function and the complementary error function. It exhibits a maximum around the unitarity, as displayed in Fig.~\ref{fig4}(a). We further define a half-way time $\tau$ as $C(t=\tau)=C(\infty)/2$ to quantify how fast the Contact evolves~\cite{quench_exp3, Sun}. The results for the BEC-BCS crossover are displayed in Fig.~\ref{fig4}(b). Surprisingly, a maximum occurs around the unitarity, indicating the evolving time is determined mainly by the difference between the final state and the initial state (see Fig.~\ref{fig4}(a)), instead of the interaction strength. 

We quantitatively investigate the damping oscillations on the BEC side to reveal the relation between oscillations and the dimer state. The relative values at the peak  $\Delta C_p=C_p-C(\infty)$ obey a universal power-law decay with the same power approximately for various interactions (see Fig.~\ref{fig4}(c)), implying the intial state plays an important role on the damping rate instead of the interaction. However, the frequencies are proportional to $a_s^{-2}$ approximately (see Fig.~\ref{fig4}(d)), which is associated to the energy difference between the dimer state and the scattering state. It will be determined by the dimer's binding energy when the energy of the scattering state can be negligible. Therefore, these oscillations of the Contact dynamics can be used as a tool to detect possible dimer states in experiments.

\section{Conclusion}

  In summary, we find a theorem showing that the continuous scale invariance of the Hamiltonian and the initial density matrix can give rise to the universal scaling dynamics. We first demonstrate it in the two-body system analytically. Then, for quantum gases quenched from the non-interacting equilibrium state, the dynamic scaling appears at large momentum region for both the unitary limit and the weak interacting limit. Remarkably, an approximate dynamic scaling with smaller scaling exponents is also found at finite momentum region, which may offer a new interpretation to the experimental observations. Our results can also guide the future exploration of the scaling dynamics in various systems. The calculated Contact dynamics exhibits several novel features such as a maximum of the half-way time around the unitary limit and the damping oscillations on the BEC side, which can be explored by the experiment in near future~\cite{quench_exp3,Zoran,Contact_exp1,Contact_exp2}.  

\section*{Acknowledgments}

  We thank Zheyu Shi, Pengfei Zhang, Ran Qi, Zhiyuan Yao, Lei Pan, Ren Zhang, and Hui Zhai for inspiring discussion. The project was supported by NSFC Grant No. 12004049, Fund of State Key Laboratory of IPOC (BUPT) (No. 600119525, 505019124).

\section*{Appendix: Proof of the theorem}
In the following, we perform the derivation of the theorem (i.e., Eq.~(\ref{nkt1}) in the main text) step by step. We first display the scaling behavior of each term separately and then put them together to give the final result. Given the quenched Hamiltonian is scale-invariant, if the wavefunction $|\phi_\eta \rangle$ is an eigenfunction of the Hamiltonian with the eigenvalue $E_\eta$( $\eta$ denotes $Nd$ quantum numbers with $N$ and $d$ being the total particle number and the dimensionality respectively), a scaled wavefunction $|\phi_{\zeta\eta} \rangle$ ($\zeta$ is a positive number) will also be an eigenfunction with the eigenvalue $E_{\zeta\eta}=\zeta^\gamma E_\eta$~\cite{Felix}. Therefore, if assuming $\phi_{\zeta\eta} (\zeta {\bf k_1},\cdots, \zeta {\bf k_N}) = A \phi_{\eta}({\bf k_1},\cdots, {\bf k_N})$  (A is a real number), one can have
\begin{widetext}
\begin{align}
 \langle  \phi_{\zeta\eta_1}  | \phi_{\zeta\eta_2} \rangle  &=  \int d{\bf k_1}\cdots d{\bf k_N} \phi_{\zeta\eta_1}^*({\bf k_1},\cdots, {\bf k_N})\phi_{\zeta\eta_2}({\bf k_1},\cdots, {\bf k_N})   \nonumber \\
   &=  \zeta^{Nd} \int d{\bf k_1}\cdots d{\bf k_N} \phi_{\zeta\eta_1}^*(\zeta{\bf k_1},\cdots, \zeta{\bf k_N})\phi_{\zeta\eta_2}(\zeta{\bf k_1},\cdots, \zeta{\bf k_N})   \nonumber \\
   &=  \zeta^{Nd} \int d{\bf k_1}\cdots d{\bf k_N} A \phi_{\eta_1}^*({\bf k_1},\cdots, {\bf k_N}) A \phi_{\eta_2}({\bf k_1},\cdots, {\bf k_N})   \nonumber \\
   &= A^2\zeta^{Nd} \langle  \phi_{\eta_1}  | \phi_{\eta_2} \rangle  \nonumber \\
    \label{tm1}
\end{align}
\end{widetext}

According to the condition $\langle  \phi_{\zeta\eta_1}  | \phi_{\zeta\eta_2} \rangle = \delta(\zeta\eta_1-\zeta\eta_2) = \zeta^{-Nd} \delta(\eta_1-\eta_2)$, one can obtain $A=\zeta^{-Nd}$ and $\langle  \phi_{\zeta\eta_1}  | \phi_{\zeta\eta_2} \rangle = \zeta^{-Nd} \langle  \phi_{\eta_1}  | \phi_{\eta_2} \rangle$. Based on this, it is straightforward to derive that
\begin{equation}
 \langle  \psi_\kappa |e^{it\hat{H}}  | \phi_\eta\rangle  =   \zeta^{Nd} \langle  \psi_{\zeta\kappa} | e^{it\zeta^{-\gamma}\hat{H}} | \phi_{\zeta\eta}\rangle. 
  \label{tm2}
\end{equation}  
For the term involved with the momentum distribution, it can be transformed as follows.
\begin{align}
   \langle  \phi_{\eta_1} |   \hat{n}_{\bf k}   | \phi_{\eta_2} \rangle &= \int d{\bf k_1}\cdots d{\bf k_N} \phi_{\eta_1}^*({\bf k_1},\cdots, {\bf k_N})  \nonumber \\ 
    &\hspace{0.5cm} \cdot \sum_{m}\delta({\bf k_m}-{\bf k})     \phi_{\eta_2}({\bf k_1},\cdots, {\bf k_N})   \nonumber \\ 
   &= \sum_{m} \int d{\bf k_m} f_{\eta_1,\eta_2}({\bf k_m})  \delta({\bf k_m}-{\bf k})      \nonumber \\ 
    &= \sum_{m} \int d{\bf k_m} \zeta^{3d} f_{\zeta\eta_1, \zeta\eta_2}(\zeta {\bf k_m})  \delta(\zeta{\bf k_m}-\zeta{\bf k})   \nonumber \\  
     &=  \zeta^{2d}  \sum_{m} \int d{\bf k_m}  f_{\zeta\eta_1, \zeta\eta_2}({\bf k_m})  \delta({\bf k_m}-\zeta{\bf k})   \nonumber \\ 
     &= \zeta^{2d} \langle  \phi_{\zeta\eta_1} |   \hat{n}_{\zeta{\bf k}}   | \phi_{\zeta\eta_2} \rangle  \nonumber \\
    \label{tm3}
\end{align}
Here, we have first integrated over all other momenta (except ${\bf k_m}$) and then regularized the scaling divergent term (see Eq.~(\ref{tm1})), to obtain $f_{\eta_1,\eta_2}({\bf k_m})=\int d{\bf k_1}\cdots d{\bf k_{m-1}}d{\bf k_{m+1}} \cdots d{\bf k_N} \phi_{\eta_1}^*({\bf k_1},\cdots, {\bf k_N}) \phi_{\eta_2}({\bf k_1},\cdots, {\bf k_N})$ in one-particle space, which satisfies $f_{\eta_1,\eta_2}({\bf k_m})=\zeta^{2d} f_{\zeta\eta_1,\zeta\eta_2}(\zeta{\bf k_m})$.

With Eq.~(\ref{tm2}) and (\ref{tm3}), one can derive the dynamic scaling of the momentum distribution as follows.
\begin{align}
   n({\bf k},t) &=\int d\kappa  w_\kappa  \langle  \psi_\kappa |e^{it\hat{H}}  \hat{n}_{\bf k}  e^{-it\hat{H}} | \psi_\kappa\rangle  \nonumber \\
    &=\int d\kappa  \int d\eta_1 \int d\eta_2 w_\kappa  \langle  \psi_\kappa |e^{it\hat{H}}  | \phi_{\eta_1} \rangle  \langle  \phi_{\eta_1} |   \hat{n}_{\bf k}   | \phi_{\eta_2} \rangle \nonumber \\
    &\hspace{0.5cm} \cdot  \langle \phi_{\eta_2} | e^{-it\hat{H}} | \psi_\kappa\rangle  \nonumber \\
    &= \zeta^{2d}\int d\kappa  w_\kappa  \langle  \psi_\kappa |e^{it\zeta^{-\gamma}\hat{H}}  \hat{n}_{\zeta{\bf k}}  e^{-it\zeta^{-\gamma}\hat{H}} | \psi_\kappa\rangle  \nonumber \\           
    &=\zeta^{2d} n(\zeta {\bf k}, t\zeta^{-\gamma}) \nonumber \\
    &\equiv \tilde{t}^{\alpha} n(\tilde{t}^{\beta} {\bf k}, t_0) \nonumber \\
    \label{nktproof}
\end{align}
Here, $w_\kappa=\zeta^{-\delta} w_{\zeta\kappa}$ with $\delta=-Nd$, which can be obtained from the relation $\int d\kappa  w_\kappa = \zeta^{Nd} \int d\kappa  w_{\zeta\kappa} = \zeta^{Nd+\delta} \int d\kappa  w_{\kappa} $. We set $\tilde{t}=t/t_0$, the scaling exponent $\beta=1/\gamma$ and $\alpha=2d/\gamma=2d\beta$. Thus, the theorem is proved.


\begin{thebibliography}{99}
\bibitem{footnote1}
In three-dimensional unitary Bose gas, the scale invariance is only approximate because of the three-body parameter. However,  experiments~\cite{quench_exp1,quench_exp2,quench_exp3} show that its contribution can be insignificant in quench dynamics.
\end{thebibliography}
\end{document}